# Enhancement of CO detection in Al doped graphene

Z. M. Ao,[†,‡] J. Yang,[‡] S. Li,[*,‡] Q. Jiang[*,†]

Key Laboratory of Automobile Materials, Ministry of Education, and Department of Materials Science and Engineering, Jilin University, Changchun 130025, China, and School of Materials Science and Engineering, The University of New South Wales, NSW 2052, Australia

**Abstract:**

A principle of enhancement CO adsorption was developed theoretically by using density functional theory through doping Al into graphene. The results show that the Al doped graphene has strong chemisorption of CO molecule by forming Al−CO bond, where CO onto intrinsic graphene remains weak physisorption. Furthermore, the enhancement of CO sensitivity in the Al doped graphene is determined by a large electrical conductivity change after adsorption, where CO absorption leads to increase of electrical conductivity upon via introducing large amount of shallow acceptor states. Therefore, this newly developed Al doped graphene would be an excellent candidate for sensing CO gas.

---

[*] To whom correspondence should be addressed. e-mail: sean.li@unsw.edu.au (S. Li), e-mail: jiangq@jlu.edu.cn (Q. Jiang).
[†] Jilin University
[‡] The University of New South Wales

Solid-state gas sensors are renowned owing their high sensitivity, robustness and a wide range of applications as well as low cost and potential for miniaturization [1]. Carbon nanotubes (CNTs) is one of the promising nanoscale molecular sensors used to detect gas molecules with fast response time and high sensitivity at room temperature [2,3,4]. The semiconducting CNTs are sensitive to variation of electrical conductivity in the presence of gas molecules at the concentration range of ppb (parts per $10^9$). However, higher sensitive sensor is desirable for the virtual applications of industrial, environmental and military monitoring.

It was reported that the detectable range and sensitivity of the single wall carbon nanotubes (SWCNTs) can be widened and enhanced substantially through either doping technology or surface engineering [4,5,6]. For example, SWCNT coated with Pb nanoparticles has high sensitivity to $H_2$ [5], $SnO_2$/SWCNTs hybrid material shows an enhanced sensitivity to $NO_2$ [6]. The high sensitivity of boron doped SWCNT to CO and $H_2O$ absorptions has been demonstrated [4]. Most recently, Al-cluster and Al doped SWCNT assembly were suggested to be promising systems for novel molecular sensors to $NH_3$ [7] and CO [8], and the B doped SWCNTs are highly sensitive to the gaseous cyanide and formaldehyde molecules [9]. However, the devices with higher sensitivity to these toxic gases are apparently required for environmental safety issues both in workplaces and residential areas, especially in some industrial and military fields.

Graphene based device may be a solution for ultra-high sensitivity gas sensor for such applications [10,11,12]. Similar to CNT, the working principle of graphene devices as gas sensors is based on the changes of their electrical conductivity induced by the surface adsorbates, which act as either donors or acceptors associated with their chemical natures and preferential adsorption sites [1,2,3,10]. Graphene is considered to be an excellent sensor material due to its following properties: (1) graphene is a single atomic layer of graphite with surface only, this can maximize the interaction between the surface dopants and adsorbates; (2) graphene has much smaller band gap energy, $E_g$, than CNT, hence, it has extremely low Johnson noise [13,14,15], therefore, a little change of carrier concentration can cause a notable variation of electrical conductivity; and (3) graphene has limit crystal defects [13,14,15,16], which ensures a low level of excess noise caused by their thermal switching [17]. In this work, we report that the sensitivity of graphene system to CO gas could be enhanced to a higher level, which exceeds by orders of magnitudes state-of-the-art sensors, through Al doping. This may provide new insight to the development of next generation gas sensors for virtual applications.

All density functional theory (DFT) calculations are performed in Dmol$^3$ code [18,19]. It is widely known that calculations limited at the local density approximation (LDA) overestimate bond energy $E_b$ and underestimate equilibrium distances [20,21]. Thus, a uniform generalized gradient approximation (GGA) with the revised Perdew-Burke-Ernzerhof (PBE) method is used as the exchange correlation function [22]. The DFT semicore pseudopotentials (DSPP) core treatment [23] is implemented for relativistic effects, which replaces core electrons by a single effective potential. To ensure that the results of the calculations are directly comparable, identical conditions are employed for all adsorption systems. The $k$-point is set to 6×6×2 for all slabs, which brings out the convergence tolerance of energy of $1.0 \times 10^{-5}$ hartree (1 hartree = 27.2114 eV), and that of maximum force of 0.002 hartree.

In the simulation, three-dimensional periodic boundary condition is taken and C–O bond length is set to $l_{C-O} = 1.13$ Å, which is consistent with experimental results [24]. For graphene, a single layer 3×3 supercell with a vacuum width of 12 Å above is constructed, which ensures that the interaction between repeated slabs in a direction normal to the surface is small enough. The variation of energetic results would be within a range of 0.1 eV if the vacuum width is expanded from 12 to 15 Å. All atoms are allowed to relax for all energy calculations. The $E_b$ between the CO gas molecule and graphene is defined as,

$$E_b = E_{CO+graphene} - (E_{graphene} + E_{CO}) \quad (1)$$

where the subscripts CO+graphene, graphene, and CO denote the adsorbed system, isolated graphene and CO molecules, respectively.

For CO adsorption on the intrinsic or Al doped graphene, there are two highly symmetric categories of adsorption configurations: one is CO molecule resides parallel to a graphene surface [Figs. 1(a)–(f)], and another is CO molecule resides perpendicular to a graphene surface [Figs. 1(g)–(l)]. These configurations are similar as that of NO on carbon nanotube [25].

To evaluate the interaction between a CO molecule and the intrinsic graphene or Al doped graphene, $E_b$ described in Eq. (1) and the binding distance, $d$, with all available configurations are calculated. Twelve available binding sites for the CO adsorbed on graphene layer are considered as initial structures. After full relaxation, no distinct structural change has been found. All of the results are displayed in Table 1. It is found that adsorption configuration shown Fig. 1(f) has the smallest $d$ value and the largest $E_b$ value among the available adsorption configurations. This indicates that the configuration shown in Fig. 1(f) is the most stable atomic arrangement with the lowest energy and the strongest interaction between CO and graphene with $E_b = 0.016$ eV and $d = 3.768$ Å which are consistent with the simulation results of $E_b = 0.014$ eV and $d = 3.740$ Å [11]. However, in this particular adsorption configuration the $E_b$ is still considered too small and $d$ too large although they are the most favorable one for adsorption, reflecting that CO undergoes weak physisorption on the intrinsic graphene. This indicates that the intrinsic graphene is insensitive to CO molecules.

When one carbon atom is substituted by Al atom in the super cell, it is found that the geometric structure of the Al doped graphene changes dramatically. Figs. 2(a) and 2(b) represent the geometries of intrinsic and Al doped graphene after relaxation. As shown in Table 2 and Fig. 2(b), the Al doping results in $l$ elongation from $l_{C-C} = 1.420$ Å to $l_{Al-C} = 1.632$ Å. This is associated with the distortion of hexagonal structures adjacent to the larger Al atom, similar to the restructuring in Al doped SWCNTs [8].

When a CO molecule is adsorbed on the Al doped graphene, which one C atom substituted by an Al atom in the super cell, there are also twelve available adsorption sites similar to the CO absorption in intrinsic graphene shown in Fig. 1. These are taken as initial configurations. After relaxation, the configuration in Fig. 2(d) has the most stable relaxed structure obtained from the initial arrangements of T−B−T, T−H−T, T (O upwards), B (O upwards) and H (O upwards) where the letters of T, B and H denote the sites of atom and CO molecule center on graphene ring, and they are defined in the caption of Fig. 1. Note that in Table 1, the deviation of $E_b$ and $l_{Al-C}$ of the four configurations above are within the error of 1%. The adsorption of CO causes a structure change in the Al doped graphene dramatically, resulting in an expansion of $l_{Al1-C2}$ from 1.632 to 1.870 Å while $l_{Al1-C4}$ elongates from 1.632 to 1.915 Å. The corresponding distance between the CO molecule and Al atom in the Al doped graphene is 1.964 Å, being much shorter than 3.767 Å in the intrinsic graphene system. Moreover, it is found that $E_b$ of CO in Al doped graphene systems are over 60 times larger than that of CO in the intrinsic graphene systems, indicating that the Al doped graphene are energetically favorable for CO adsorption.

Furthermore, to investigate the changes of electronic structures in graphenes caused by the physi- or chemisorption of CO molecule, the net electron transfer ($Q$) from either the intrinsic or the Al doped graphene to the polar CO molecules are calculated by Mulliken population analysis, where $Q$ is defined as the charge variation caused by the CO absorption. As listed in Table 2, $Q = 0.027$ $e$ in the Al doped graphene is almost one order larger than 0.003 $e$ in the intrinsic graphene. This supports the notion that the Al doping influences the electronic properties of graphene substantially. This can also be verified by the difference of electronic densities between the intrinsic and Al doped graphenes with and without the CO adsorption. In Fig. 3, the red and blue regions represent the areas of electron accumulation and the electron loss, respectively. Fig. 3(a) indicates the bond in the intrinsic graphene is of covalent nature because the preferential electron accumulation sites are mainly located within the bond rather than heavily centered on a particular atom. However, the Al doping modified the electron density by inducing the different electron affinities for Al and C atoms but the whole structure remains covalent in nature [Fig. 3(b)]. Physisorption of CO on the intrinsic graphene does not alter the electron distribution for both CO molecule and graphene, implying the weak bonding characteristics. It is discernable that electronic polarization is induced by the preferential accumulation of electrons on O in CO molecules [Fig. 3(c)]. As distinct from the CO absorption on the intrinsic graphene, the chemisorption of CO on Al doped graphene leads to significant electron transfer from the graphene to CO molecule [Fig. 3(d)]. In this case, the electrons not only accumulate on the O atom but also on the C atom of the molecule bond with the doped Al atom. The final position of Al atom in the chemisorbed CO−Al−graphene complex is thus a direct consequence of the maximized degree of $sp^3$ orbital hybridization with neighboring C atoms from both the graphene layer and CO molecule. This is evidential because the red lobes around C atoms in Fig. 3(d) are both pointing towards Al atom.

To further determine the effects of CO absorption on electrical conductivity, the electronic densities of state (DOS) for the both systems with and without the absorption are calculated. As shown in Figs. 4(a) and (b), the Al doping in graphene enhances its electrical conductivity by shifting the highest DOS peak to just below the Fermi level $E_f$, which also leads to decrease reduction of $E_g$. This indicates that the doping Al atom induces shallow acceptor states in graphene like B atom in SWCNs, thus enhancing its extrinsic conductivity [4]. When the CO molecule adsorbed on the intrinsic and doped graphene surfaces, the total DOSs are shown in Figs. 4(c) and 4(d). In the intrinsic graphene, the DOS of CO−graphene system near $E_\text{f}$ have no distinct change, and the conductivity change is barely observable. It implies that the intrinsic graphene would not be an ideal CO gas sensor. However, for the Al doped graphene with the most stable chemisorbed CO configuration [Fig. 2(d)], not only the highest DOS peak shifts over the $E_\text{f}$, but also the DOS value increases dramatically. This results in a gap closure [Fig. 4(d)], suggesting that an extra number of shallow acceptor states are introduced when the Al doped graphene interacts with the highly polar CO molecule. As a result, the chemisorbed CO on the Al doped graphene will give rise to a large increase in the electrical conductivity of the doped graphene layer. By detecting the conductivity change of the Al doped graphene systems before and after the adsorption of CO, the presence of this toxic molecule can be detected sensitively. Therefore, the Al doped graphene is a promising sensor material for detecting CO molecules. However, desorption of CO molecule from the Al doped graphene is difficult due to the strong bonding of Al−CO [25]. This can be solved by applying an electric field $F$ to reactivate the sensor materials [26].

In conclusion, the adsorptions of CO molecules on the intrinsic and Al doped graphenes are investigated using DFT calculation. It is found that CO molecules are only weakly adsorbed onto the intrinsic graphene with small binding energy value and large distance between the CO molecules and graphene. The electronic structure and electrical conductivity of the intrinsic graphene have a limit change caused by the adsorption of CO molecules. However, the CO molecule has strong interaction with the Al doped graphene, forming an Al−CO bond that introduces a large amount of shallow acceptor states into the system. In this case, the remarkable variation of the electrical conductivity is induced by the CO adsorption, possessing an excellent characteristic of high sensitivity for CO gas detection.

**Acknowledgment**

This work was financially supported by National Key Basic Research and Development Program (Grant No. 2004CB619301), ″985 Project″ of Jilin University and Australia Research Council Discovery Program DP0665539.

Table 1. Summary of results for CO adsorption on intrinsic graphene and Al doped graphene on different adsorption sites. The meaning of T, B and H are given in the caption of Fig. 1.

| Initial binding configuration | | Intrinsic graphene | | Al doped graphene | |
|---|---|---|---|---|---|
| | | $E_b$ (eV) | $d$ (Å)[a] | $E_b$ (eV) | $l$ (Å)[b] |
| CO//graphene | T–B–T | -0.011 | 3.839 | -4.978 | 1.964 |
| | T–H–T | -0.012 | 3.805 | -4.973 | 1.968 |
| | H–T–H | -0.014 | 3.826 | -4.613 | 3.755[a] |
| | H–B–H | -0.009 | 3.857 | -4.599 | 3.814[a] |
| | B(C atom)–T–H | -0.011 | 3.855 | -4.609 | 3.800[a] |
| | B(O atom)–T–H | -0.016 | 3.768 | -4.616 | 3.821[a] |
| CO⊥graphene | T(O upwards) | -0.007 | 3.938 | -4.979 | 1.961 |
| | B(O upwards) | -0.007 | 3.935 | -4.978 | 1.964 |
| | H(O upwards) | -0.003 | 3.982 | -4.975 | 1.965 |
| | T(C upwards) | -0.004 | 3.952 | -4.629 | 3.781[a] |
| | B(C upwards) | -0.003 | 3.981 | -4.607 | 3.783[a] |
| | H(C upwards) | -0.005 | 3.942 | -4.609 | 3.457[a] |

a. Binding distance between CO gas molecule and graphene layer.
b. Bond length of Al and C atom in CO gas molecule.

Table 2. Some structure parameters of intrinsic graphene and Al doped graphene before and after adsorption of CO molecule.

| System | Configuration | Bond | Bond length $l$ (Å) | $Q$ ($e$)[a] |
|---|---|---|---|---|
| Intrinsic graphene | Fig. 2(a) | C1–C2 | 1.420 | |
| | | C1–C3 | 1.420 | |
| | | C1–C4 | 1.420 | |
| | Fig. 2(c) | C1–C2 | 1.420 | 0.003 |
| | | C1–C3 | 1.421 | |
| | | C1–C4 | 1.421 | |
| Al doped graphene | Fig. 2(b) | Al1–C2 | 1.632 | |
| | | Al1–C3 | 1.632 | |
| | | Al1–C4 | 1.632 | |
| | Fig. 2(d) | Al1–C2 | 1.870 | 0.027 |
| | | Al1–C3 | 1.910 | |
| | | Al1–C4 | 1.915 | |

a. Electrons transferred from the graphene layer to CO molecule. $e$ denotes one electronic charge.

**Captions:**

Fig. 1. Twelve available binding sites for CO adsorbed on intrinsic graphene (top and below images show the top and side view, respectively). (a) T–B–T, (b) T–H–T, (c) H–T–H, (d) H–B–H, (e) B(C atom)–T–H, (f) B(O atom)–T–H, (g) T (O atom upward), (h) B (O atom upward), (i) H (O atom upward), (j) T (C atom upward), (k) B (C atom upward), (l) H (C atom upward). T, B and H denote top site of C atoms, bridge site of C–C bond and hollow site of carbon hexagon, respectively. Gray, pink and red spheres are denoted as C, Al and O atoms, respectively. In Figs. 2 and 4, spheres have the same meaning as that in Fig. 1.

Fig. 2. Atomic configurations of intrinsic graphene and Al doped graphene before and after adsorption of CO gas molecule where one Al atom dopes in site 1, and sites 2, 3 and 4 are C atoms near the doped Al atom. (c) and (d) are the preferred configurations after CO adsorption for intrinsic graphene and Al doped graphene, respectively.

Fig. 3. Images of the electronic density difference for intrinsic graphene (a), Al doped graphene (b), CO−graphene system with preferred configuration (c) and CO−Al doped graphene system with preferred configuration (d). The red region shows the electron accumulation, while the blue region shows the electron loss.

Fig. 4. Electronic density of state (DOS) of intrinsic graphene (a), Al doped graphene (b), CO−graphene system with preferred configuration (c), and CO−Al doped graphene system with preferred configuration (d).

Fig. 1

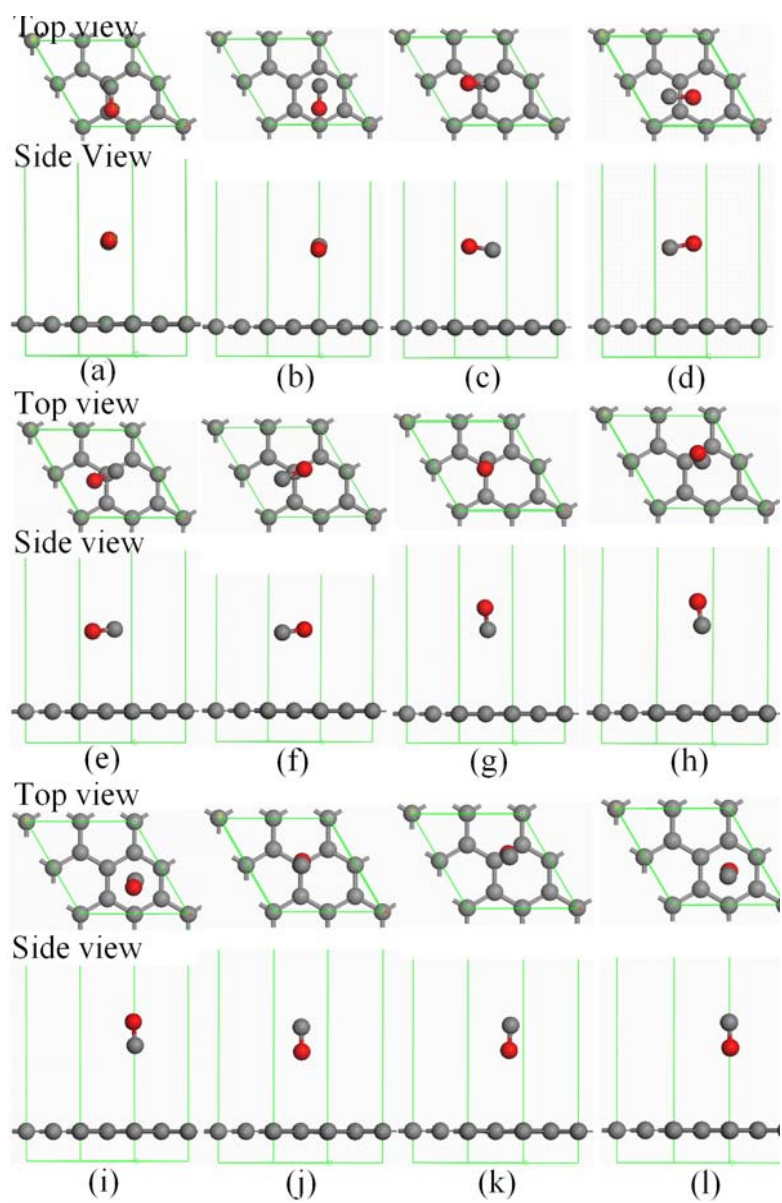

Fig. 2

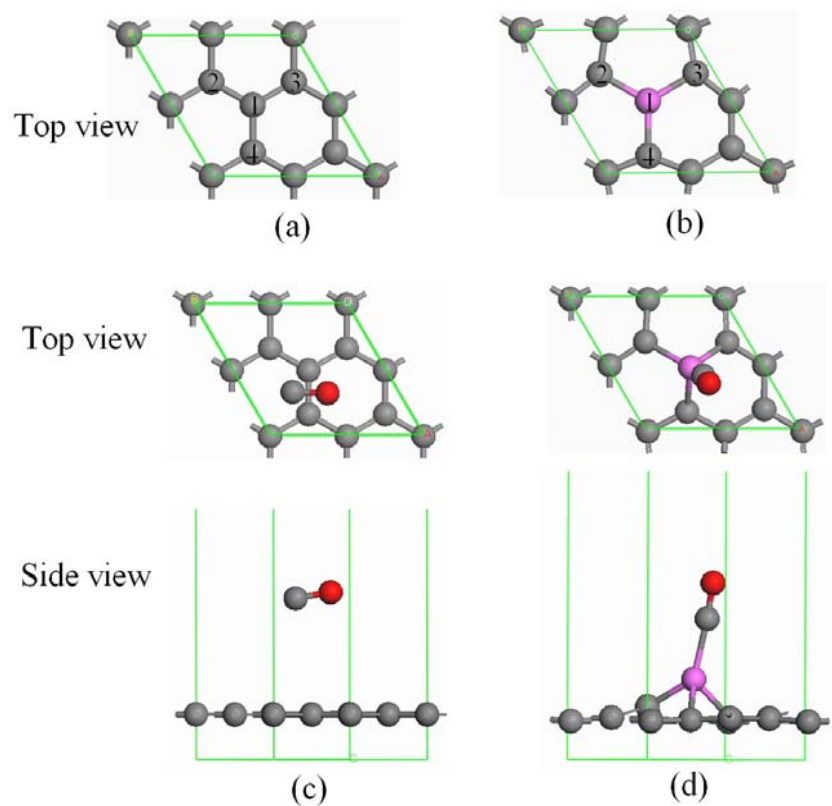

Fig. 3

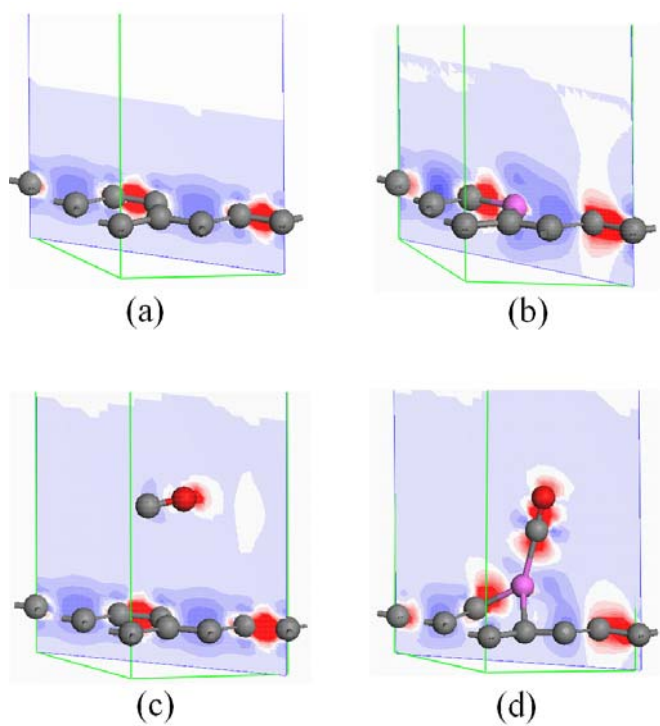

Fig. 4

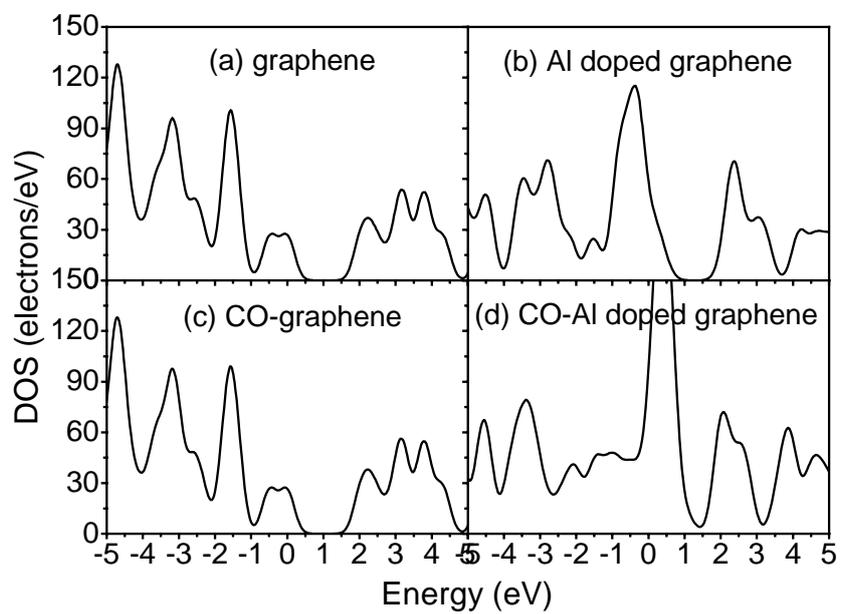